\documentclass[12pt]{article}
\usepackage{a4,amsmath,epsfig}
\usepackage{times,euler,euscript}
\renewcommand{\baselinestretch}{1.1}
\newcommand{\Section}[1]{Section~\ref{#1}}

\newcommand{\Figure}[1]{Fig.\ref{#1}}
\newcommand{\Equation}[1]{Eq.(\ref{#1})}
\newcommand{\ie}{{\it i.e.}}
\newcommand{\parni}{{\tt PARNI}}
\newcommand{\vegas}{{\tt VEGAS}}
\newcommand{\ndim}{D}
\newcommand{\cube}{I_{\ndim}}
\newcommand{\xdat}{x}
\newcommand{\idat}{i}
\newcommand{\ndat}{n}
\newcommand{\sequence}[3]{(#1_{i})_{i=#2}^{#3}}
\newcommand{\Exp}{\mathsf{E}}
\newcommand{\Var}{\mathsf{V}}
\newcommand{\intcube}{\int_{\cube}\!\!}
\newcommand{\dxdat}{d^{\ndim}\xdat}
\newcommand{\funcf}{f}
\newcommand{\funcg}{g}
\newcommand{\ich}{k}
\newcommand{\jch}{l}
\newcommand{\nch}{m}
\newcommand{\wch}{w}
\newcommand{\gch}{g}
\newcommand{\rch}{A}
\newcommand{\fch}[2]{F^{(#1)}_{#2}}
\newcommand{\sch}[2]{S^{(#1)}_{#2}}
\newcommand{\funcs}{s}
\newcommand{\ydat}{y}
\begin{document}
\begin{center}
%
\begin{flushright} \mbox{\normalsize IFJPAN-IV-2007-13} \end{flushright}

\vspace{2\baselineskip}
{\bf\Huge PARNI for importance sampling}\\\vspace{0.25\baselineskip}
{\bf\Huge and density estimation%
\footnote{Supported in part  by the EU  RTN European Programme,
MRTN-CT-2006-035505 (HEPTOOLS, Tools and Precision Calculations for
Physics Discoveries at  Colliders)
and by the Polish Ministry of
Scientific Research and Information Technology grant No 153/6 PR UE/2007/7
2007-2010.}
}

\vspace{2\baselineskip}
{\Large Andr\'e van Hameren%
}

\vspace{0.25\baselineskip}
{\it\large IFJ-PAN, Krak\'ow, Poland}

\vspace{0.25\baselineskip}
{\tt\large Andre.Hameren@ifj.edu.pl}

\vspace{0.25\baselineskip}
{\large\today}

\renewcommand{\baselinestretch}{1}
\vspace{2\baselineskip}
{\bf Abstract}\\\vspace{0.5\baselineskip}
\parbox{0.8\linewidth}{\small\hspace{15pt}%
We present an aid for importance sampling in Monte Carlo integration, which is of the general-purpose type in the sense that it in principle deals with any quadratically integrable integrand on a unit hyper-cube of arbitrary dimension.
In contrast to most existing systems of this type, it does not ask for the integrand as an input variable, but provides a number of routines which can be plugged into a given Monte Carlo program in order to improve its efficiency ``on the fly'' while running.
Due to the nature of its design, it can also be used for density estimation, \ie, for the analysis of data points coming from an external source.\\[4ex]
}
\end{center}
\vspace{\baselineskip}

\section{Introduction}
%
Numerical integration often is the only solution to integration problems encountered in scientific research.
If the space over which the integral has to be performed, the {\em integration space\/}, has many dimensions, the Monte Carlo method usually appears to be the only feasible option.
It has the advantage that it works for any quadratically integrable function, but the disadvantage that the convergence rate may be low, \ie, that many function evaluations may be needed to obtain a result to acceptable accuracy.
Via the method of {\em importance sampling\/}, however, it is possible to translate information about the integrand into an enhancement of the rate of convergence, in principle up to the point where only one function evaluation is needed to complete the integral.
In the latter case, the integration problem has essentially been solved analytically.

Obviously, information about the integrand is obtained during the Monte Carlo integration process itself through the evaluation of the integrand at the integration points.
Importance sampling based solely on this information is called {\em adaptive\/} importance sampling.
It is the last resort to improve the rate of convergence, possibly only applied to a sub-region of the integration space.
In particular, it can be useful for so called {\em general-purpose\/} integration systems which are supposed to be able to deal with many different integrands.
Several such systems have been developed and are being used extensively in the field of elementary particle phenomenology \cite{Lepage:1977sw,Kawabata:1995th,Jadach:2002kn,Hahn:2005pf,Nason:2007vt}.
Most of these use the evaluated integration points to adapt the treatment of sub-spaces in which they are generated during the integration process.

The ability to apply importance sampling successfully implies the availability of a probability density which has a shape comparable to the shape of the absolute value of the integrand.
An attempt to use importance sampling is essentially the search and construction of such a probability density.
Consequently, given such a density, one can generate points in the integration space which are distributed following the shape of the absolute value of the integrand, or at least one has a tool to increase the efficiency of such generation process using straightforward methods like rejection.
This is an important issue if one wants to perform Monte Carlo {\em simulations\/}, and in fact, many of the aforementioned general-purpose systems have been designed also with this goal in mind.

One technical item most of the mentioned general-purpose systems have in common is that they are {\em integrators\/}, possibly providing the probability density to be used for simulation after the integration process.
More specifically, the integrand is an input variable, and the integration process is performed by the system itself.
Sometimes, however, one has his own Monte Carlo program, and one would like to improve parts of it with adaptive importance sampling ``on the fly'', without having to turn it into a function that can be used as the input in an initial ``integration phase''.
In this write-up, we present the program \parni, which has been designed with exactly this in mind.
It consists, apart of a trivial initialization, essentially of three routines which can be called inside a Monte Carlo program; one for the generation of integration points, one to return their weights, and one to collect the evaluated function values at those points.
The collection routine builds the probability density with which the integration points are generated, and works independently from the generation routine.
This means it can take points from another generation process, and estimate the density following which they are generated.
In particular, the function evaluation may be absent and all points may have weight one, so that \parni\ can serve as a pure density estimator.

The outline of the paper is as follows.
In \Section{SecAdaptive} an issue concerning the possible overestimation of integration errors when using adaptive importance sampling is addressed.
\Section{SecAlgorithm} explains how the algorithms work on which \parni\ is based.
In \Section{SecBinary} a technical detail concerning its architecture is addressed.
\Section{SecUse} explains how the program is used, and \Section{SecExamples} gives some examples.
\Section{SecSummary} finally contains the summary.
%

\section{Adaptive importance sampling\label{SecAdaptive}}
In this section we address an issue in the application of adaptive importance sampling in Monte Carlo integration concerning possible overestimates of the integration error. 
First of all, we assume that the integration/generation problem has been formulated such that the integration space is a hyper-cube $\cube=[0,1)^{\ndim}$ of a certain dimension $\ndim$.
Monte Carlo integration is based on the fact that for a quadratically integrable function $\funcf$ and a sequence of points $\sequence{\xdat}{1}{\ndat}$ in the hyper-cube distributed independently following a probability density $\funcg$, 
the distribution of the average of the ratio $\funcf/\funcg$ over the sequence converges, following the Central Limit Theorem, to a Gaussian distribution with expectation value
%
\begin{equation}
\Exp\bigg(
\frac{1}{\ndat}
\sum_{\idat=1}^{\ndat}\frac{\funcf(\xdat_{\idat})}{\funcg(\xdat_{\idat})}
\bigg)
=
\intcube\funcf(\xdat)\,\dxdat
~,
\end{equation}
%
and variance
%
\begin{equation}
\Var\bigg(
\frac{1}{\ndat}
\sum_{\idat=1}^{\ndat}\frac{\funcf(\xdat_{\idat})}{\funcg(\xdat_{\idat})}
\bigg)
=
\frac{1}{\ndat}
\bigg[
\intcube\frac{\funcf(\xdat)^2}{\funcg(\xdat)}\,\dxdat
-\bigg(\intcube\funcf(\xdat)\,\dxdat\bigg)^2\,
\bigg]
~.
\label{Eq163}
\end{equation}
%
Therefore, the average can be interpreted as an estimate of the integral, with a possible integration error given by the square root of its variance.
The variance can be estimated by
%
\begin{equation}
\frac{1}{\ndat-1}
\bigg[
\frac{1}{\ndat}\sum_{\idat=1}^{\ndat}\frac{\funcf(\xdat_{\idat})^2}{\funcg(\xdat_{\idat})^2}
-
\bigg(\frac{1}{\ndat}\sum_{\idat=1}^{\ndat}\frac{\funcf(\xdat_{\idat})}{\funcg(\xdat_{\idat})}\bigg)^2\,
\bigg]
~,
\end{equation}
%
and approaches zero for large values of $\ndat$.
The only restrictions on the density $\funcg$ are that it is non-zero on the support of $\funcf$, and that there is an algorithm available to generate the sequence of points distributed following this density which is not too complex.
Of course, the evaluation of the density at the integration points should be feasible, or say, should not be much more complex than the evaluation of the integrand.
A first candidate for the density $\funcg$ is unity on the hyper-cube.
Importance sampling is the attempt to construct a density $\funcg$ such that the variance (\ref{Eq163}) is, for a given value of $\ndat$, as small as possible.

In adaptive importance sampling, the evaluated integration points are used to update the density during the integration process.
In this case, the {\em L\'evi Central Limit Theorem\/}, which is the basis for the version of Monte Carlo integration presented above, does not hold anymore since it applies only in case the points are {\em independently identically\/} distributed.
Theoretically, this does not need to be problem since the {\em Martingale Central Limit Theorem\/} may still apply \cite{vanHameren:2001ng}.
In practice, however, it means that the integration error is possibly unnecessarily overestimated.
To see how this happens, assume that a batch of $\ndat$ points has been generated with density $\funcg_{1}$, and another batch with density $\funcg_{2}$ which possibly depends on the first batch of data.
The straightforward Monte Carlo estimate of the integral would be
%
\begin{equation}
Y
=
\frac{X_{1}+X_{2}}{2}
\quad\textrm{with}\quad
X_{1}
=
\frac{1}{\ndat}\sum_{i=1}^{\ndat}
\frac{\funcf(\xdat_{\idat})}{\funcg_{1}(\xdat_{\idat})}
\quad\textrm{and}\quad
X_{2}
=
\frac{1}{\ndat}\sum_{i=\ndat+1}^{2\ndat}
\frac{\funcf(\xdat_{\idat})}{\funcg_{2}(\xdat_{\idat})}
~.
\end{equation}
%
The variance of the estimate is given by%
\footnote{The correlation $\Exp(X_{1}X_{2})-\Exp(X_{1})\Exp(X_{2})$ is zero. This is an indication for the martingale structure of a sequence of Monte Carlo estimates using adaptive importance sampling.}
%
\begin{equation}
\Var(Y)
=
\frac{\Var(X_{1})+\Var(X_{2})}{4}
~.
\end{equation}
%
If the process by which $\funcg_{2}$ is updated using the first batch of data is efficient enough, it may happen that $\Var(X_{2})<\frac{1}{3}\Var(X_{1})$, and the result for the integral using both batches may be worse than the result using only the second batch.

The straightforward solution to this problem is to introduce an ``optimization phase'' in the integration process, in which points are generated and the integrand is evaluated to adapt the density, but do not contribute to the final estimate of the integral.
The integral is estimated in a second ``integration phase'', in which the optimized density is not changed anymore, and the L\'evi Central Limit Theorem applies.

The optimal solution would be to weigh the contributions with the inverse of the variance, \ie, to take
%
\begin{equation}
Y
=
\frac{\Var(X_{1})\Var(X_{2})}{\Var(X_{1})+\Var(X_{2})}\,
\bigg(\frac{X_{1}}{\Var(X_{1})} + \frac{X_{2}}{\Var(X_{2})}\bigg)
~.
\end{equation}
%
This choice gives the minimum variance for $Y$, however, the variances have to be estimated too, and replacing them by their estimates in the above formula would lead to a biased estimator for the integral of $\funcf$.

A third option is to weigh contributions of equal sized batches of integration points, which are generated with the same density, with pre-determined weights.
\parni\ in particular generates batches of constant size with the same density, and uses the information from a whole batch at once to update the density, with which the next batch is generated.
Weighing the contributions with the order in the sequence, such that the first batch gets a relative weight equal to $1$, the second batch gets a relative weight equal to $2$ etc., appears to be a safe choice in the sense that the contribution of the early batches is relatively low and the problem sketched above is avoided.

\section{The algorithm\label{SecAlgorithm}}
The main ingredient of the algorithm \parni\ uses to build the probability density is to divide the hyper-cube into sub-regions on which the density is constant, in a way similar to other systems, in particular the one in \cite{Jadach:2002kn}.
The second ingredient is to consider the probability density a weighted sum of probability densities restricted to those sub-regions.
This is essentially an application of the {\em multi-channel\/} method \cite{Kleiss:1994qy}.%
\footnote{Notice the difference with the application of the multi-channel method in \cite{Ohl:1998jn}, where each channel consists of a probability density on the full hyper-cube.}
It excellently suits the task of updating the probability density ``on the fly'', by collecting batches of integration points of constant size during the generation of which the density is not changed.
The disadvantage is that it requires extra memory during the computation, since the weights of all the sub-densities have to be stored, something which is avoided for example in \cite{Jadach:2002kn}.

To be more specific, at a given stage during the Monte Carlo computation, the density is given by
%
\begin{equation}
\funcg(\xdat)
=
\sum_{\ich=1}^{\nch}\wch_{\ich}\gch_{\ich}(\xdat)
~,
\end{equation}
%
where each $\gch_{\ich}$ is the constant probability density on a sub-region $\rch_{\ich}$ of the hyper-cube, and the weights $\wch_{\ich}$ are positive and sum up to $1$
%
\begin{equation}
\sum_{\ich=1}^{\nch}\wch_{\ich} = 1
~.
\end{equation}
%
The regions are non-overlapping hyper-rectangles, and their union is the hyper-cube:
%
\begin{equation}
\forall_{\ich\neq\jch}~\rch_{\ich}\cap\rch_{\jch} = \emptyset
\quad\textrm{and}\quad
\bigcup_{\ich=1}^{\nch}\rch_{\ich} = \cube
~.
\end{equation}
%
A triple $(\rch_{\ich},\gch_{\ich},\wch_{\ich})$ will be referred to as {\em channel} number $\ich$.

The creation of new channels is incorporated such that it aims at equidistribution, \ie, it aims at moving the values of the channel weights towards the average $1/\nch$.
Given a batch of evaluated integration points and a method how to update the weights, one of the updated weights, say $\wch_{\ich}$, will be the largest.
Hyper-rectangle $\rch_{\ich}$ is divided into two equal sized pieces, and $\gch_{\ich}$ becomes a weighted sum of two probability densities.
The weight of each of the these new densities in the full sum of densities simply becomes half of the original updated weight.
The division of the original rectangle happens such that it is cut in the middle perpendicular to the dimension along which it has the longest edges.
If it has several dimensions for which the edges are of equal length and the longest, one of them is chosen at random.
This way, the hyper-rectangles tend to have the shape of hyper-cubes.
After one division, we have a new set of $\nch'=\nch+1$ channels.
The division process is be repeated until the weight efficiency
%
\begin{equation}
\frac{1}{\nch'\,\max_{\ich}\wch_{\ich}}
\end{equation}
%
does not increase anymore (notice that $\nch'$ increases while $\max_{\ich}\wch_{\ich}$ decreases with each division).

We have reserved a few options how to update the weights before the division process.
In order to achieve variance minimization, the relative weights should be taken proportional to%
\footnote{Realize that $\gch_{\ich}$ is a {\em probability density\/}, so it is the indicator function of $\rch_{\ich}$ divided by the volume of $\rch_{\ich}$.}
%
\begin{equation}
\wch_{\ich}
\propto
\sqrt{\mathrm{vol}(\rch_{\ich})^2\intcube\funcf(x)^2\funcg_{\ich}(x)\,\dxdat}
~,
\label{EqWeightUpdate2}
\end{equation}
%
where $\mathrm{vol}(\rch_{\ich})$ is the volume of sub-region $\rch_{\ich}$.
The integral is estimated by $\fch{2}{\ich}/\fch{0}{\ich}$, where
%
\begin{equation}
\fch{p}{\ich} = \sum_{\xdat_{\idat}\in\rch_{\ich}}\funcf(\xdat_{\idat})^p
~.
\end{equation}
%
We do not calculate the numbers $\fch{p}{\ich}$ per batch, but calculate them incrementally during the whole adaptation process, by dividing them by $2$ if the corresponding channel is divided in two pieces during in the creation of channels.

Variance minimization might not lead to the optimal density to be used for simulation purposes; the density with which the simulation is most efficient.
For this, a more point-wise recovery of the integrand may be desirable.
Better results may be expected by putting the relative weights proportional to 
%
\begin{equation}
\wch_{\ich}
\propto
\mathrm{vol}(\rch_{\ich})\intcube\funcf(x)\funcg_{\ich}(x)\,\dxdat
~.
\label{EqWeightUpdate1}
\end{equation}
%
%
The integral is estimated by $\fch{1}{\ich}/\fch{0}{\ich}$.

In case \parni\ is used for the task of density estimation, implying that only the full weight coming with each data point is supplied, the channel weights are put proportional to
%
\begin{equation}
\wch_{\ich}
\propto
\sch{1}{\ich}
\end{equation}
%
where
%
\begin{equation}
\sch{p}{\ich} = \sum_{\xdat_{\idat}\in\rch_{\ich}}\funcs_{\idat}^p
\end{equation}
%
and $\funcs_{\idat}$ is the weight coming with $\xdat_{\idat}$.
This way essentially a multi-dimensional histogram with non-equal sized bins is being built.

With the creation of more and more channels, more and more memory is needed, which may not be available anymore at some point.
\parni\ provides the option to set a maximum to the number of channels.
If this number is reached, \parni\ will continue to divide channels, but it will also start to merge channels in order to keep the total number close to the maximum set.
Merging is performed with the channels which came from the same parent channel during the division process, and have the smallest sum of weights of all such pairs.
The original parent hyper-rectangle is restored, and it gets a channel weight which is the sum of the weights of the pair of daughter channels.
Also the values of the quantities $\fch{p}{\ich}$ and $\sch{p}{\ich}$ are obtained by adding those of the daughters together.

\section{Binary-tree structure\label{SecBinary}}
One technical detail which is worth mentioning because it highly contributes to the computational efficiency of the program is that the structure of sub-spaces is, like in \cite{Jadach:2002kn}, organized in a binary-tree structure.
So for every hyper-rectangle, the program keeps track of the parent it was created from by the division process, and the daughters created by its own division.
This means that the program keeps track of twice as many hyper-rectangles as there are channels, but this disadvantage is fully compensated by the increase in computational efficiency.
For example, in order to evaluate the built density at a given point in the hyper-cube, the hyper-rectangle has to be found in which the point is situated.
Because of the binary-tree structure this can be done very quickly with a binary search.
Also for the generation of a point in the hyper-cube following the built density a binary-tree search is used.
The channel weights are put in a row on the unit interval, and a random number is thrown in.
Now the interval, corresponding to the weight in the row, has to be found in which this random number exactly fell.
This interval then corresponds to the channel delivering the next integration point.
%

\section{Use of the program\label{SecUse}}
The program is written in {\tt Fortran77\/} using long names, underscores in names, \verb|enddo| and \verb|do while| statements.
It has been designed such that an arbitrary number of instances of the algorithm can work in parallel, dealing with completely different integrands living in different numbers of dimensions.
Before \parni\ can be used, some integer type global parameters have to be set.
This happens in the header-file \verb|avh_parni.h|.
There is \verb|avh_parni_size|, which is a measure of the total amount of memory one is willing to spend on all instances of \parni\ together.
The amount consists of the equivalent of this parameter times $4$ double precisions plus $6$ integers.
Then there is \verb|avh_parni_ncopy|, setting the maximal number of instances of \parni, and finally there is \verb|avh_parni_dim|, setting the largest dimension an instance of \parni\ may be ordered to deal with. 

Next, the user of the program has to implement the contents of the the routine
\begin{verbatim}
subroutine avh_parni_random(rho ,nn)
\end{verbatim}
which is supposed to generate double precision arrays \verb|rho| of arbitrary length \verb|nn| consisting of uniformly distributed (pseudo) random numbers.
In practice one line with a call to an external routine of this type will suffice.

An instance of \parni\ has to be initialized before the Monte Carlo loop with
\begin{verbatim}
call avh_parni_init(ID ,itask ,ndimen ,nbatch ,nchmax)
\end{verbatim}
All variables are input and integers, and the first one, \verb|ID|, is the id of the instance of \parni.
This number should be larger than $0$ and not larger than the value of \verb|avh_parni_ncopy|.
The second one, \verb|itask|, specifies the way the channel weights are updated.
The choice \verb|itask=1| corresponds to \Equation{EqWeightUpdate1} and \verb|itask=2| corresponds to \Equation{EqWeightUpdate2}.
For the task of density estimation, one has to put \verb|itask=11|.
The third input variable, \verb|ndimen|, specifies the number of dimensions of the hyper-cube.
The fourth input variable, \verb|nbatch|, specifies the size of the batches of data points to be collected before a new adaptation step is performed.
The larger this number, the more conservative the algorithm and the less channels will be created for a given total number of data points.
From experience we find that this number should best be of the order of the square-root of the total number of data points that is expected to be encountered.
The fifth input variable, \verb|nchmax|, is the maximal number of channels one wishes to reach.
If its value is put to $0$, the number of channels will grow indefinitely.

Inside the Monte Carlo loop, an integration point distributed following the density built so far is generated by
\begin{verbatim}
call avh_parni_generate(ID ,xx)
\end{verbatim}
The first variable, \verb|ID|, is input and is the id of the instance of \parni.
The second variable, \verb|xx|, is output consisting of a double precision array of length \verb|ndimen|, where the latter is the number given on input at the initialization of this instance of \parni.
This integration point is stored together with its weight, and these data are returned by
\begin{verbatim}
call avh_parni_weight(ID ,ww,xx)
\end{verbatim}
Here, the weight \verb|ww| and the point \verb|xx| are output.
One can obtain the value of the built probability density at any point inside the hyper-cube with the function
\begin{verbatim}
avh_parni_density(ID ,xx)
\end{verbatim}
Here, \verb|xx| is input.
The weight coming with an integration point is given by $1$ divided by the number obtained by evaluating this function at the generated point.
A data point is collected for the adaptation of the density with
\begin{verbatim}
call avh_parni_adapt(ID ,value ,xx)
\end{verbatim}
All variables are input.
The third one, \verb|xx|, is the data point in the hyper-cube, and the second one, \verb|value|, is the full weight coming with this point, so including the weight from the generation of \parni\ itself in the case of importance sampling.

Besides the necessary routines described above, there are some diagnostic routines available.
With
\begin{verbatim}
call avh_parni_marg(ID ,nunit ,label)
\end{verbatim}
for each dimension a file is created with two columns of data corresponding to the marginal density in that dimension.
In other words, it gives the density obtained when all other dimensions are ``integrated out''.
The integer input variable \verb|nunit| refers to the unit the file is written to, and \verb|label| is, if it is chosen to be positive, an extra label to the file name.
If, for example, \verb|ID=13| and \verb|label=25|, then the file name with the marginal density of dimension number \verb|2| is \verb|PARNI13d2_25|.

In case the integration space is $2$-dimensional, the density can be visualized with the file created by
\begin{verbatim}
call avh_parni_plot(ID ,nunit ,label)
\end{verbatim}
In the example above, the name of the file name will be \verb|PARNI13p_25|.
It is in a format to be used by {\tt gnuplot} \cite{gnuplot}:
\begin{verbatim}
gnuplot> plot 'PARNIp13_25' w l
\end{verbatim}
will visualize the $2$-dimensional structure of rectangles, whereas
\begin{verbatim}
gnuplot> splot 'PARNIp13_25' w l
\end{verbatim}
will visualize a full $3$-dimensional picture of the density.

The number of channels which finally has been reached is send to standard output by
\begin{verbatim}
call avh_parni_result(ID)
\end{verbatim}
Also an integration estimate with error estimate are printed, but the latter may be an overestimation.

Finally we want to remark that the program is distributed over $3$ source files with a total length of less than $1500$ lines including comments, and that the names of all subroutines, functions and common blocks start with \verb|avh_parni|.
%

\section{Some examples\label{SecExamples}}
%
A typical problem in a general purpose Monte Carlo program for elementary particle phenomenology is that certain variables which have to be generated may be distributed very differently for different processes.
An extreme example would be a distribution consisting of a sharp delta-like spike with an unknown position.
In order to simulate this situation, we use \parni\ to integrate a function consisting of a truncated Cauchy-density
%
\begin{equation}
\funcf(\xdat)
=
\frac{N(\xdat_{0},\varepsilon)}{(\xdat-\xdat_{0})^2 + \varepsilon^2}
~,
\label{Eq636}
\end{equation}
%
where $N(\xdat_{0},\varepsilon)$ is the normalization such that the correct integral is equal to $1$.
We take $\xdat_{0}=0.6$ and $\varepsilon=10^{-5}$, so that relative to the unit interval the density becomes a sharp delta-like spike.
\begin{figure}
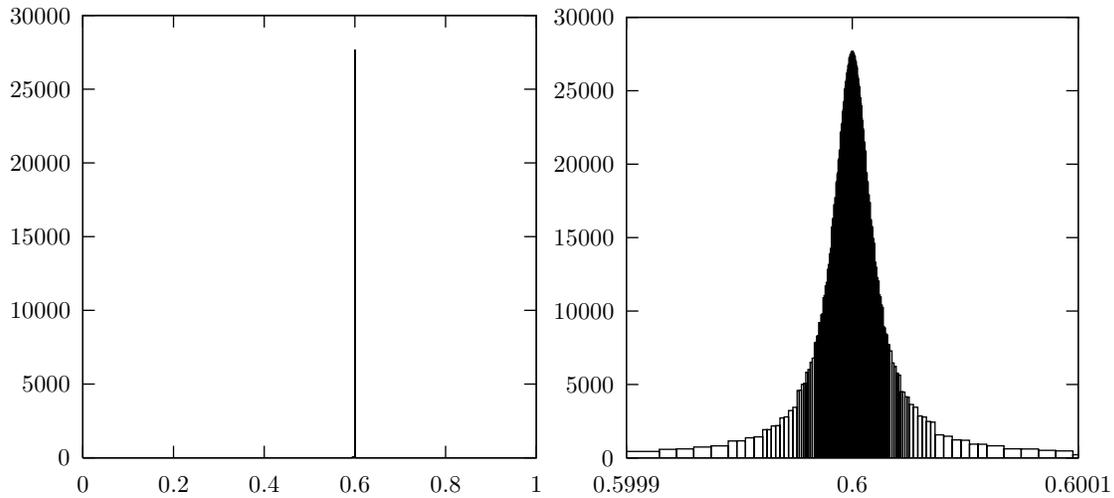

\begin{center}
\epsfig{file=fig7.eps,width=0.44\linewidth}
\epsfig{file=fig8.eps,width=0.46\linewidth}
\caption{The density built by \parni\ during the integration of the function in \Equation{Eq636} with $\varepsilon=10^{-5}$. The picture on the right is a zoom-in.}
\label{Fig4}
\end{center}
\end{figure}
In \Figure{Fig4} we show the density built by \parni\ after $10^4$ integration points applied to the adaptation in batches of $10^2$.
We did not restrict the number of channels, and \parni\ created $202$ of them.
To give a quantitative measure of the result: integration of the integrand with importance sampling using this density (so after its creation, not during) is done with a crude efficiency%
\footnote{The average weight divided by the maximal weight in the Monte Carlo process.}
of about $23\%$. 
Without any importance sampling, the Monte Carlo integration of this integrand would be performed with a crude efficiency of $0.0037\%$.

The program \vegas\ \cite{Lepage:1977sw} works very well for multi-dimensional integration problems for which the integrand is factorisable over the dimensions, \ie, when the integrand can be seen as a product of integrands with each of them depending on only one of the dimensions.
In the second example, we visualize why it is very important to use this knowledge about the integrand when available.
We choose an integrand which is the product of truncated Cauchy-densities
%
\begin{equation}
\funcf(\xdat,\ydat)
=
\frac{N(\xdat_{0},a)}{(\xdat-\xdat_{0})^2 + a^2}
\cdot
\frac{N(\ydat_{0},b)}{(\ydat-\ydat_{0})^2 + b^2}
~,
\label{Eq637}
\end{equation}
%
with $(\xdat_{0},a)=(0.6,0.02)$  and $(\ydat_{0},b)=(0.33,0.04)$. 
Now we attack the integration problem in two different ways.
In the first approach, we let one instance of \parni\ deal with the full $2$-dimensional problem at once.
A rudimentary Monte Carlo program in this approach would look like
\begin{verbatim}
      call avh_parni_init(1,itask,2,nbatch,nchmax) ! ndimen=2
      result = 0d0
      do iev=1,nev
        call avh_parni_generate(1,x)
        weight = integrand(x(1),x(2))
     &         / avh_parni_density(1,x)
        call avh_parni_adapt(1,weight,x)
        result = result+weight
      enddo
      result = result/nev
\end{verbatim}
In the second approach, we let two instances of \parni\ deal with the problem, each of them with one of the dimensions.
A rudimentary Monte Carlo program in this approach would look like
\begin{verbatim}
      call avh_parni_init(1,itask1,1,nbatch1,nchmax1) ! ndimen=1
      call avh_parni_init(2,itask2,1,nbatch2,nchmax2) ! ndimen=1
      result = 0d0
      do iev=1,nev
        call avh_parni_generate(1,x1)
        call avh_parni_generate(2,x2)
        weight = integrand(x1,x2)
     &         / avh_parni_density(1,x1)
     &         / avh_parni_density(2,x2)
        call avh_parni_adapt(1,weight,x1)
        call avh_parni_adapt(2,weight,x2)
        result = result+weight
      enddo
      result = result/nev
\end{verbatim}
In order to compare the two approaches in a fair way, we demand that the sum of the number of channels used by the two instances in the latter approach should be the same as the number of channels used by the one instance in the former approach.
\begin{figure}
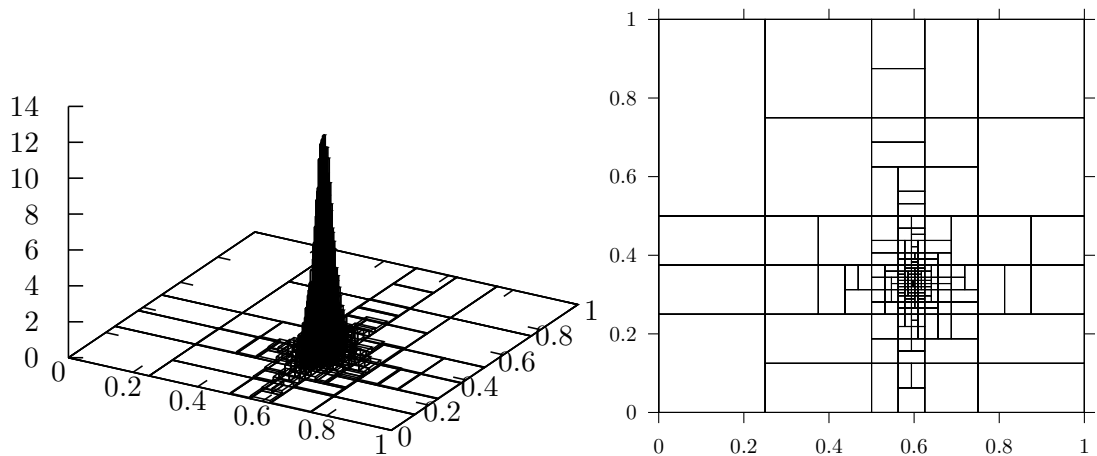

\begin{center}
\epsfig{file=fig1.eps,width=0.5\linewidth}
\epsfig{file=fig2.eps,width=0.4\linewidth}
\caption{Density (left) and structure of rectangles (right) built during the integration of the integrand in \Equation{Eq637}.}
\label{Fig1}
\end{center}
\end{figure}
\Figure{Fig1} depicts the density built using $10^5$ integration points in batches of $316$ in the first approach.
The number of channels is $200$.
The picture on the right is the top-view of the picture on the left, and shows the structure of rectangles.
\begin{figure}
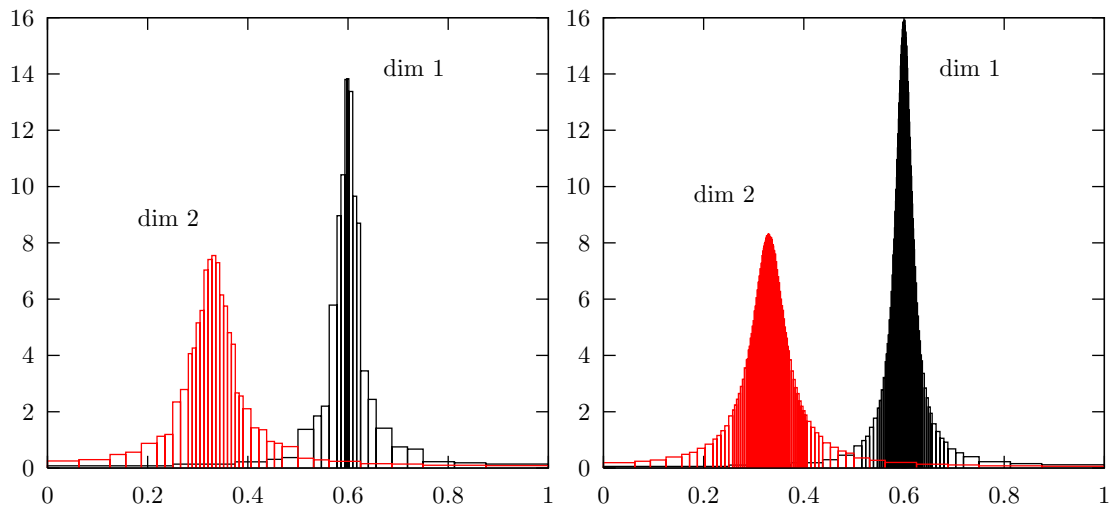

\begin{center}
\epsfig{file=fig3.eps,width=0.45\linewidth}
\epsfig{file=fig4.eps,width=0.45\linewidth}
\caption{Marginal densities obtained with one instance of \parni\ dealing with the full problem (left) and with the \vegas-approach (right), using the same total number of channels.}
\label{Fig2}
\end{center}
\end{figure}
The picture on the left of \Figure{Fig2} shows the marginal densities in the two dimensions for this case.
The picture on the right, on the other hand, shows the density built by each of the instances of \parni\ in the second approach.
In this picture, the densities look much nicer because more ``bins'' are available, namely $100$ for each of them, whereas in the picture on the left this number is only of the order of $\sqrt{200}$.
Since we are dealing with a factorisable integrand, we know that the marginal densities are sufficient to capture all information about the integrand, and we can thus confirm visually that the second, \vegas-like, approach works much better.
This is also shown more quantitatively by the crude efficiency of the integration of the integrand using the densities (again after their creation, not during): $66\%$ for the \vegas-approach against $15\%$ for the other approach.

When the integrand is not factorisable, \vegas\ does not work so well, and for example \cite{Jadach:2002kn} has particularly been designed to deal with this problem.
Also \parni\ can deal with non-factorisable integrands, and as an example, we give some results for a $2$-dimensional integration problem in which the support of the integrand is concentrated around a circle.
It is proportional to
%
\begin{equation}
\funcf(\xdat,\ydat)
\propto
\exp\bigg(-\frac{(r-c)^2}{d^2}\bigg)
\quad\textrm{with}\quad
r = \sqrt{(\xdat-a)^2+(\ydat-b)^2}
~,
\label{Eq741}
\end{equation}
%
and we put $a=0.57$, $b=0.62$, $c=0.3$ and $d=0.01$.
\begin{figure}
\begin{center}
\epsfig{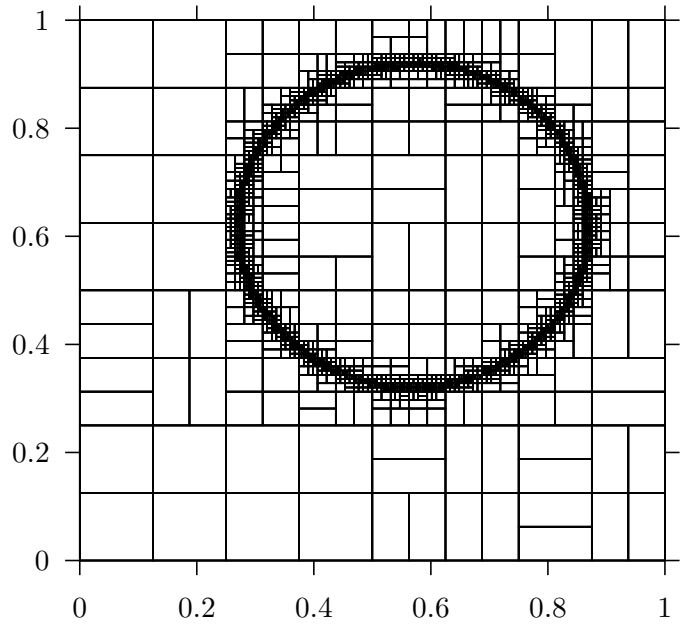}
\caption{Structure of rectangles built during the integration of the integrand in \Equation{Eq741}.}
\label{Fig3}
\end{center}
\end{figure}
The structure of rectangles built by \parni\ after $10^6$ integration points in batches of $10^3$ is shown in \Figure{Fig3}.
We did not put a restriction on the number of channels to be created, and \parni\ created $1819$ of them.
If we try to deal with this integrand using the \vegas-like approach, around $2800$ channels are created for either dimension.
Still, the integration is less efficient: with the same number of integration points, the \vegas-like approach reaches an estimate of the relative error of $0.24\%$, whereas the approach with one instance dealing with the full problem reaches an estimated $0.081\%$.
In comparison, without any importance sampling, one reaches an estimated relative error of $0.43\%$.
%

\section{Summary\label{SecSummary}}
We presented the program \parni, a practical aid for importance sampling in Monte Carlo integration.
It adapts automatically to integrands on the unit hyper-cube of in principle any dimension, and can therefore be considered to be of the general-purpose type.
However, it does not ask for the integrand as an input variable, but provides a number of routines which should be incorporated into an existing Monte Carlo program, so that the optimization happens ``on the fly'' while the Monte Carlo is running.
The generation of integration points and the adaptation of the probability density following which they are generated happen independently, and the program can also be used only to adapt the probability density using data from another source.
In other words, it can also be used as a pure density estimator.
%



\begin{appendix}

\end{appendix}
%
%
\end{document}